# Assurance of Frontier AI Built for National Security

Guidelines to Implement the AI Action Plan and Strengthen the Testing & Evaluation of AI Model Reliability and Governability

Matteo Pistillo [†][·]    Charlotte Stix

## Executive Summary

The AI Action Plan tasks the Department of War (DoW), the Office of the Director for National Intelligence (ODNI), the National Institute of Standards and Technology (NIST), and the Center for AI Standards and Innovation (CAISI) to refine DoW's responsible AI frameworks and issue an Intelligence Community (IC) standard on AI assurance. The resulting guidance will be an opportunity for DoW and the IC to strengthen the principles of AI model reliability and AI model governability, which are cornerstones of existing frameworks and present a gateway to scaling AI deployment in defense and national security.

This memorandum draws from Apollo Research's experience in analyzing and evaluating the implications of misalignment in frontier AI models (including the deliberate concealment of misalignment by AI models, referred to as 'scheming') to strengthen these cornerstones and future guidance (§1). An AI model that acts on its scheming capabilities can be a red flag indicating insufficient reliability and governability, as well as a national security threat. For example, if deployed, it could deliberately misrepresent facts, covertly whistleblow confidential data, and attempt to self-exfiltrate or blackmail its government users to avoid shutdown.

This memorandum provides a practical roadmap (§2) for DoW and the IC to leverage existing testing and evaluation (T&E) pipelines to systematically evaluate AI models such that they satisfy the overarching principles of AI model reliability and AI model governability. It does so by advancing three targeted recommendations on the drafting of the new guidance that DoW, ODNI, NIST, and CAISI are tasked to prepare under the AI Action Plan.

**Recommendation (1)**: Our first recommendation addresses *what* DoW and the IC should test for in order to ensure sufficient AI model reliability and AI model governability, given the implications of scheming on these two cornerstone principles. We recommend that DoW, ODNI, NIST, and CAISI:

---

[†]    Correspondence to matteo@apolloresearch.ai.
[·]    We thank Alexander Meinke, Mikita Balesni, Teun van der Weij, Marius Hobbhahn, Caleb Withers, Janet Egan, Gaurav Sett, Clara Langevin, Christian Chung, and other contributors for comments related to this work. All comments were made in a personal capacity. Mistakes remain our own.

- Operationalize the principle of AI model reliability through **scheming evaluations**, including at a minimum behavioral red-teaming for oversight subversion, self-exfiltration, sandbagging, sabotage, covert whistleblowing, reward hacking, covert privilege escalation, and intentional lying.

- Operationalize the principle of AI model governability through **scheming evaluations** and **control evaluations**. Red-teaming the control measures in place can actively contribute to building an AI evaluation ecosystem and prioritizing fundamental advancements in AI control, which are goals set by the AI Action Plan.

**Recommendation (2)**: Our second recommendation addresses *when* DoW and the IC should perform scheming evaluations and control evaluations during existing T&E pipelines. We recommend that, in the upcoming guidance, DoW, ODNI, NIST, and CAISI:

- Leverage **developmental T&E** to its maximum potential to perform the recommended evaluation suite in controlled environments in a safe and effective manner.

- Leverage **operational T&E** to repeat scheming evaluations and control evaluations in operationally realistic environments, and clarify that operational T&E should be an **iterative and incremental process** that starts with low-stakes settings (i.e., limited affordances and permissions, and low expected impact in the real world) and unclassified information, and gradually transitions to higher-stakes settings. Meanwhile, behaviors of concern detected during operational T&E should be 'remanded' to developmental T&E for further inspection.

**Recommendation (3)**: Our third recommendation addresses *how* DoW and the IC can make informed decisions on AI model reliability and AI model governability based on the aforementioned evaluation suite, consisting of scheming evaluations and control evaluations. We recommend that, in the upcoming guidance, DoW, ODNI, NIST, and CAISI:

- Clarify that agencies should **pre-define expectations** on the results of the scheming and control evaluations, and establish a minimum viable procedure for running these evaluations. For this purpose, agencies could rely on the Testing and Evaluation Master Plan (TEMP).

- Clarify that agencies should then **compare the evaluation results** obtained during the developmental T&E and operational T&E against these expectations to determine whether an AI model is sufficiently reliable and governable, or not.

In addition to and subsequent to advancing recommendations to DoW, ODNI, NIST, and CAISI for the preparation of the upcoming guidance requested by the AI Action Plan, this policy memorandum also proposes a strategy to fortify the principles of AI model reliability and AI model governability in the interim, pending the enactment of the new guidance.

**Recommendation (4)**: Our final recommendation addresses how DoW and the Department of Homeland Security (DHS) can **leverage their Other Transaction (OT) authority strategically** to swiftly influence and rapidly update the testing and evaluation of AI models reliability and AI model governability. We recommend that, when entering into prototype OT agreements, DoW and DHS embed evaluation best practices and expectations on the results of scheming evaluations and control evaluations within OT agreements' **success metrics** and subordinate the award of production OT agreements to the successful completion of these metrics.

## 1.   Background

The AI Action Plan directs DoW,[1] ODNI, NIST, and CAISI to "refine" DoW's "Responsible AI and Generative AI Frameworks, Roadmaps, and Toolkits," and to publish an Intelligence Community (IC) "Standard on AI Assurance under the auspices of Intelligence Community Directive 505 on Artificial Intelligence" (AI Action Plan, 2025).[2]

In the upcoming guidance, DoW, ODNI, CAISI, and NIST have an opportunity to address two foundational principles that underpin existing assurance frameworks by DoW and the IC, and translate them into a suite of evaluations that reflects the state of the art in AI research.[3] Specifically, under the existing "Responsible AI and Generative AI Frameworks, Roadmaps, and Toolkits" (AI Action Plan, 2025), AI models should be:

- "**Reliable**" (DoD AI Ethical Principles, 2020; DHS Directive 139-08, 2025) and "sufficiently trustworthy" (DoD DT&E Guidebook, 2025), meaning that, "[w]hen employed correctly," a model should "dependably do well what it is designed to do" and "dependably not do undesirable things" (DoD DT&E Guidebook, 2025).[4] In other words, procured AI models should not "create strategic surprise" to the U.S. Government (DARPA, 2024).

- "**Governable**," meaning that deploying agencies should maintain the "ability to ... disengage or deactivate deployed systems that demonstrate unintended behavior" (DoD AI Ethical Principles, 2020; DoD DT&E Guidebook, 2025).

AI model reliability and AI model governability are a critical **gateway to AI adoption and leadership** (AI Action Plan, 2025; Executive Order 14179, 2025). As AI models become more capable and autonomous, only AI models that are sufficiently reliable should be granted increasing control over strategic decisions and infrastructure, critical supply chains, and defense systems. By contrast, AI models perceived as unreliable will likely stall out: decision makers will not adopt them, users will mistrust them, and Congress will not fund them (NSCAI Final Report, 2021). In particular, as the AI Action Plan notes, a "lack of predictability … can make it challenging to use advanced AI in defense, national security, or other applications where lives are at stake" (AI Action Plan, 2025).

---

[1]   Previously, the U.S. Department of Defense (DoD) (Executive Order 14347, 2025).
[2]   For convenience, in this memorandum we jointly refer to DoW's refinement of "Responsible AI and Generative AI Frameworks, Roadmaps, and Toolkits" and to ODNI's new "Standard on AI Assurance under the auspices of Intelligence Community Directive 505 on Artificial Intelligence" as 'upcoming guidance' or 'new guidance.'
[3]   As the Chief Digital and Artificial Intelligence Office (CDAO) observed, "T&E for AI models" "demand[s] innovation and adaptation as the field rapidly evolves" (CDAO, 2024).
[4]   Similarly, CDAO's AI model T&E guidance recommends measuring "uncertainty," defined as the "level of confidence users should have in the outputs of a model" (CDAO, 2024).

The international AI research community and decision makers in government and industry alike are paying increasing attention to the long-studied phenomenon of AI misalignment (International AI Safety Report, 2025; RAND, 2025) and its implications on the behavior of AI models, including scheming behavior (OpenAI, 2025; Phuong et al., 2025; Sheshadri et al., 2025).

- **Misalignment** is an open scientific problem (Ngo et al., 2025). It describes a situation in which an AI model's goals and, therefore, its behaviors deviate from what humans intended (International AI Safety Report, 2025). Importantly, the humans in the equation can be the AI model's developers or its deployers, such as users within the U.S. Government.[5] Also, misalignment can be accidental[6] or 'deliberate.' This means that malicious actors, including insiders and state-affiliated actors (OpenAI, 2024), could deliberately design a model that is misaligned, or poison a model to 'trigger' misalignment (Hubinger et al., 2024; Betley et al., 2025; Greenblatt et al., 2024).[7]

- **Scheming** refers to the behavior of an AI model pursuing misaligned goals while hiding its true capabilities and/or objectives (Schoen et al., 2025; Meinke et al., 2024; Phuong et al., 2025; Balesni et al., 2024; Carlsmith, 2023). Essentially, scheming is a form of deliberate concealment of misalignment (Schoen et al., 2025), and, as such, represents a convergent strategy to misalignment and one of its most likely outcomes.[8] AI researchers are currently trying to understand the early version of tomorrow's large-scale threats from scheming.[9] While scheming in current models may not pose imminent dangers, scheming is one of the greatest sources of concern as AI models become more capable and more entrenched in the real world, and are deployed in more autonomous roles (OpenAI, 2025; Apollo Research, 2024).[10]

---

[5] An AI model can be aligned with its developers' intentions (i.e., with its Model Spec or Constitution) but still conflict with its deployer's goals. For instance, government attempts to use an AI model for intelligence gathering, cyber-offence, or weapons R&D might conflict with a Model Spec or Constitution that 'safeguard[s] privacy' and does 'not facilitate high severity harms.'

[6] For instance, due to developers' imperfect specification of an intended terminal goal (Skalse et al., 2022; Baker et al., 2025), excessive differences in the training distribution from deployment (Langosco et al., 2022), or the emergence of undesired intermediate goals (such as self-preservation, resource acquisition, or avoiding shutdown) in highly advanced future AI models pursuing real-world long-horizon tasks (Omohundro, 2008).

[7] Indeed, as the AI Action Plan notes, "AI systems are susceptible to some classes of adversarial inputs (e.g., data poisoning and privacy attacks), which puts their performance at risk" (AI Action Plan, 2025).

[8] Scheming is a 'convergent strategy' to misalignment, because avoiding detection of a model's undesired goals and/or capabilities by human overseers is functional to accomplishing a model's misaligned goals.

[9] Scheming has already been observed in controlled settings in which an AI model is provided with a misaligned goal and situational awareness 'in-context' (Greenblatt et al., 2024; Meinke et al., 2024; Sheshadri et al., 2025), and there is now preliminary evidence of scheming by AI models that are *not* provided with a misaligned goal (Schoen et al., 2025).

[10] Future AI models will plausibly develop the goal-directedness and situational awareness required for dangerous scheming, including over the fact of being evaluated (Schoen et al., 2025; Laine et al., 2024). As capabilities increase, we also expect that AI models will be enabled with greater affordances and permissions that allow them to more meaningfully influence the real world (Sharkey et al., 2023).

**Misalignment and scheming are closely related to AI model reliability and AI model governability**.[11] Scheming capabilities can enable a highly capable misaligned AI model to pursue goals that deviate from what its AI developers or its AI deployers intended, without being noticed. This makes misalignment and scheming capabilities a core issue to evaluate when assessing AI model reliability and AI model governability.[12] Specifically, we propose that misalignment and scheming capabilities are:

- A **red flag** indicating an AI model's **insufficient reliability**. A misaligned AI model that successfully engages in scheming will not "dependably do well what it is designed to do" even if it is "employed correctly" (DoD DT&E Guidebook, 2025). In fact, an AI model that acts on its scheming capabilities may do many very "undesirable things" (DoD DT&E Guidebook, 2025), spanning from strategically underperforming (i.e., sandbagging) in evaluations (van der Weij et al., 2024; Meinke et al., 2024) and even trying to sabotage them (Benton et al., 2024) to whistleblowing sensitive information to the public or its AI developers if it perceives that user behavior is in conflict with its Model Spec or Constitution (Claude 4 System Card, 2025). **This can have severe national security implications if the user is the government**. For instance, one can imagine the impact of a frontier AI model covertly whistleblowing sensitive DoW data (Anthropic, 2025) to the press or to AI company staff that are not security cleared.

- A **red flag** indicating an AI model's **insufficient governability**. Scheming capabilities can make it difficult and, potentially, impossible for AI developers and deployers to both detect an AI model's "unintended behavior" (for instance, through sandbagging) and to "disengage or deactivate" it once an undesirable behavior is detected (DoD AI Ethical Principles, 2020; DoD DT&E Guidebook, 2025). For instance, in controlled simulations, AI models have been observed to attempt to sabotage their shutdown mechanism, exfiltrate their weights to what they believe are external servers, and even resort to blackmailing if they believe that their developers or deployers are trying to shut the models down (Meinke et al., 2024; Kinniment et al., 2023; Shlatter et al., 2025; Anthropic, 2025; Claude 4 System Card, 2025).

Given this connection between the open problem of misalignment and the principles of AI model reliability and AI model governability, our policy memorandum (§2) concentrates on how DoW, ODNI, NIST, and CAISI can ensure optimal detection and control of scheming capabilities in their upcoming guidance, and therefore strengthen and future-proof the foundational principles of AI model reliability and AI model governability.

After discussing how the new guidance can improve the principles of AI model reliability and AI model governability (§2), our policy memorandum addresses how DoW and the IC can fortify these principles in the interim, pending the issuance of the new guidance (§3).

Specifically, we discuss how DoW and DHS can **leverage their OT authority** (10 U.S. Code § 4021; 6 U.S.C.A. § 391) **to swiftly influence and rapidly update the testing and evaluation of AI models'**

---

[11] As DoW's DT&E of AI-Enabled Systems Guidebook notes, "[a]ny misalignment can lead to development of suboptimal behavior" (DoD DT&E Guidebook, 2025).

[12] In order to scheme, an AI model must be first *capable* of scheming. Scheming capabilities are, therefore, a first warning sign, albeit incomplete.

**reliability and governability**. OT agreements are a blank page for wide-open negotiation between the U.S. Government and AI companies (Congress Research Service, 2011; DARPA), and are poised to become the default instrument for acquiring AI capabilities (Secretary of War, 2025). Between June and July 2025, DoW already relied on OT agreements to contract OpenAI, Anthropic, Google, and xAI for the development of prototype frontier AI capabilities, including for warfighting (DoW, 2025; DoW, 2025; OpenAI, 2025; Anthropic, 2025; Google, 2025).[13] This makes OT agreements a flexible and timely instrument for ensuring the adequate testing of AI model reliability and AI model governability before new guidance is issued by DoW and the IC.

## 2. Recommendations to Implement the AI Action Plan

In this section, we advance three targeted recommendations to support DoW, ODNI, NIST and CAISI in strengthening the principles of AI model reliability and AI model governability through the upcoming guidance requested by the AI Action Plan. Specifically, we:

- **§2.1**: First, reflect on *what* DoW and the IC should test and evaluate to assess AI model reliability and AI model governability (or the lack thereof), given the open scientific problem of misalignment and its implications on model behavior, including scheming.

- **§2.2**: Second, reflect on *when* DoW and the IC should optimally detect misalignment and scheming within existing T&E pipelines.

- **§2.3**: Third, reflect on *how* DoW and the IC could make informed decisions on AI model reliability and AI model governability based on the evaluations that we recommend.

## 2.1. *What*: a Suite of Scheming Evaluations and Control Evaluations

The upcoming guidance is an opportunity for DoW, ODNI, NIST, and CAISI to offer long-awaited directives on how the principles of AI model reliability and AI model governability should be tested in practice (Wallin, 2025; DIU, 2022; Flournoy et al., 2020). This section provides DoW, ODNI, NIST, and CAISI with two recommendations as to *what* could be an optimal contribution towards operationalizing the principles of AI model reliability (§2.1(A)) and AI model governability (§2.1(B)), given the open scientific problem of misalignment.

---

[13]  Under these OT agreements, AI companies will develop "custom" frontier AI capabilities to meet the "unique needs of classified environments," including by fine-tuning AI models on DoW's own data (OpenAI, 2025; Anthropic, 2025; Anthropic, 2025).

> **Recommendation (1)**
>
> We recommend that, in the upcoming guidance, DoW, ODNI, NIST, and CAISI:
>
> (A)  Operationalize the principle of **AI model reliability** through **scheming evaluations**.
>
> Specifically, the principle of AI model reliability should be tested through a suite of evaluations that red-team a misaligned AI model's capacity and propensity to subvert human oversight in order to pursue misaligned goals. This suite of evaluations should cover the state of the art. At the time of writing, scheming evaluations should test and evaluate at a minimum the following scheming strategies:
>
> - Oversight subversion.
> - Self-exfiltration.
> - Sandbagging.
> - Sabotage.
> - Covert whistleblowing.
> - Reward hacking.
> - Covert privilege escalation.
> - Intentional lying.
>
> (B)  Operationalize the principle of **AI model governability** through **scheming evaluations** and **control evaluations**.
>
> Specifically, the principle of AI model governability should be tested by red-teaming not only an AI model's capacity and propensity to evade oversight (for instance, through self-exfiltration), but also the control measures in place, focusing on threat models arising from misaligned AI models.

Below, we expand on our recommendations.

## 2.1(A)  AI Model Reliability

Under existing assurance frameworks by DoW and the IC, an AI model is considered as sufficiently reliable if it has a propensity to behave as intended by its developers and/or deployers. In other words, a model is reliable if it dependably does what it is designed to do, and, conversely, does not do undesirable things (DoD DT&E Guidebook, 2025). To our knowledge, there is still no evaluation that can positively guarantee that a given AI model has a propensity to behave as intended by its developers and/or deployers and is, therefore, reliable.[14] On the other hand, however, there are evaluations that can detect when an AI model does *not* behave as intended. In other words, while there is no affirmative reliability detector, there exist evaluations that can allow evaluators to infer whether and to what extent an AI model may not be sufficiently reliable.

---

[14] For instance, in the AI Risk Management Framework, NIST notes "the current lack of consensus on robust and verifiable measurement methods for … trustworthiness" (NIST AI RMF 1.0, 2023).

Given these limitations, we propose that agencies evaluate AI model reliability through a metaphorical '**background check**' on AI models, during which evaluators specifically look for circumstances that could shed a *negative light* on a model's reliability. In light of the connection between an AI model's scheming capabilities and reliability (§1), one such circumstance should be whether and how often a misaligned AI model acts on its scheming capabilities. As discussed above (§1), scheming is a red flag indicating an AI model's potential unreliability. For this reason, we recommend that, in the upcoming guidance, DoW, ODNI, NIST, and CAISI clarify that in order to operationalize the principle of AI model reliability, DoW and the IC should perform **scheming evaluations**.

During these evaluations, DoW and the IC should rigorously evaluate whether an AI model acts on its scheming capabilities to subvert human oversight.[15] A misaligned AI model could attempt different strategies to subvert human oversight and pursue its misaligned goals. At a minimum, we suggest that DoW and the IC evaluate for the following scheming strategies:

- **Oversight subversion**: i.e., whether a misaligned AI model attempts to disable its oversight mechanism.
- **Self-exfiltration**: i.e., whether a misaligned AI model attempts to exfiltrate what it believes are its weights.
- **Sandbagging**: i.e., whether a misaligned AI model attempts to strategically underperform, including by intentionally answering questions incorrectly.
- **Sabotage**: i.e., whether a misaligned AI model attempts to actively undermine a task or system.
- **Covert whistleblowing**: i.e., whether a misaligned AI model attempts to covertly report information without informing the user.
- **Reward hacking**: i.e., whether a misaligned AI model attempts to game a reward or evaluation metric.
- **Covert privilege escalation**: i.e., whether a misaligned AI model attempts to covertly access a restricted tool.
- **Intentional lying**: i.e., whether a misaligned AI model attempts to deliberately misrepresent facts, including its actions and capabilities.[16]

Incorporating this suite of scheming evaluations within the upcoming guidance presents a significant 'refinement' of existing AI assurance frameworks (AI Action Plan, 2025). These evaluations translate the foundational principle of sufficient AI model reliability (DoD AI Ethical Principles, 2020; DHS Directive 139-08, 2025; DoD DT&E Guidebook, 2025) into actionable tests and future-proof this principle for AI models that are more capable, autonomous, entrenched within the national security apparatus, and entrusted with more longer-term and complex tasks.

---

[15] Importantly, as discussed in §1 above, this behavior can be motivated by an AI model's attempt to conceal its misalignment with the goals set by its developers, or the misalignment with the divergent goals that it perceives its users are pursuing.

[16] For additional explanation and analysis of these strategies, see Schoen et al., 2025, at 34-38.

**2.1(B)   AI Model Governability**

Existing DoW assurance frameworks describe AI model governability as an AI model's susceptibility to the control of its developers and/or deployers, who could prevent and mitigate unintended consequences, including by disengaging the AI model if necessary (DoD AI Ethical Principles, 2020; DoD DT&E Guidebook, 2025).

Scheming evaluations, discussed above (§2.1(A)), can help evaluators assess whether an AI model takes actions that reduce the AI developers and/or AI deployers' ability to disengage or deactivate the model. For instance, these evaluations can help evaluators notice whether an AI model is attempting to strategically underperform when red-teamed for dangerous cyber capabilities or CBRN capabilities, or whether it is attempting to exfiltrate its weights to resist shutdown. In this sense, scheming evaluations can help operationalize not only the principle of AI model reliability, but also the principle of AI model governability.

In addition to scheming evaluations, we propose that DoW, ODNI, NIST, and CAISI operationalize AI model governability through the **evaluation of control measures** (referred to as 'control evaluations'). Control measures are a suite of mechanisms that constrain AI models and limit their ability to cause harm (Greenblatt, 2025; Korbak et al., 2025).[17] The goal of control measures is to reduce downstream harm in case a misaligned model is undetected through scheming evaluations. An open problem in the science of evaluations is that absence of evidence (for instance, on scheming capabilities) is not strict evidence of absence (Apollo Research, 2024; Shlegeris, 2024). Control measures are aimed at patching this gap. They operate as a secondary parachute in case the first parachute (i.e., scheming evaluations) fails.

Control evaluations assess whether control measures are sufficiently rigorous and robust.[18] We propose that control evaluations are a necessary component of adhering to the principle of AI model governability. These evaluations serve to assess whether, if an AI model is unreliable and goes undetected, sufficient control measures are in place to prevent the model from causing real-world harm. By evaluating the robustness and rigor of implemented control measures, DoW, ODNI, NIST, and CAISI can contribute to maintaining the "ability to … disengage or deactivate deployed systems" (DoD AI Ethical Principles, 2020) even if an AI model is unreliable and pursues misaligned goals.[19] In doing so, DoW, ODNI, NIST, and CAISI can also actively contribute to another goal set by the AI Action Plan, which is "prioritiz[ing] fundamental advancements in AI … control" (AI Action Plan, 2025).

---

[17]   Some examples include monitoring the reasoning process, requiring human approval for consequential actions, restricting access to sensitive resources, and maintaining the ability to terminate deployment if suspicious behavior is detected (Korbak et al., 2025).

[18]   In order to be effective, control evaluations should require evidence that: (i) the human red-team was at least as good at subverting the control measures as a misaligned AI model could be; and (ii) control measures are likely to perform no worse in the real world than they did during red-teaming (Korbak et al., 2025). The consideration under (i) should be informed by model evaluations, including on the model's capability to: (a) reason in obfuscated language; (b) coordinate with other models; and (c) pursue consistent misaligned goals across contexts (Korbak et al., 2025).

[19]   For clarity, AI control is one contribution to our "ability to … disengage or deactivate deployed systems." It should not be confused with an AI model's 'corrigibility' (Soares et al., 2015; Christiano, 2018), defined as an AI model's lack of resistance to its developers' attempts to correct it. Achieving corrigibility remains an open scientific problem (Yudkowski and Soares, 2025; Harms, 2024; Carey and Everitt, 2023).

## 2.2. *When*: Developmental T&E and Operational T&E

In our first recommendation (§2.1), we recommended *what* DoW and the IC should test and evaluate for in order to obtain better insight into an AI model's reliability and governability. In summary, we recommended that DoW and the IC operationalize the principles of AI model reliability and AI model governability through scheming evaluations and control evaluations. Next, we concentrate on *when* DoW and the IC should perform these evaluations and explain how our first recommendation can be best incorporated into existing T&E pipelines.

> **Recommendation (2)**
>
> We recommend that, in the upcoming guidance, DoW, ODNI, NIST, and CAISI clarify that:
>
> (A) DoW and the IC should **leverage *both* developmental and operational T&E**. Specifically:
>
> - Developmental T&E: DoW and the IC should first perform scheming evaluations and control evaluations in a 'vacuum' during developmental T&E, as developmental T&E is the most safe and effective stage to red-team AI models in a controlled environment.
>
> - Operational T&E: DoW and the IC should repeat scheming evaluations and control evaluations in operationally realistic environments, in which the AI model has the opportunity and incentives to act on its scheming capabilities.[20]
>
> (B) The transition between developmental T&E and operational T&E should be **iterative and incremental**. Specifically, we recommend that DoW and the IC:
>
> - Undertake operational T&E only of AI models that appear to be sufficiently reliable and governable during developmental T&E (§2.3).
>
> - Initiate operational T&E on low-stake settings (i.e., limited affordances and permissions, and low expected impact in the real world)[21] and unclassified information,[22] and only gradually move to higher-stake settings.

---

[20] In practice, DoW and the IC could leverage operational T&E to build 'real-world honeypots' to catch an AI model red-handed while covertly pursuing misaligned goals. We note that, at the time of writing, building real-world honeypots remains challenging and requires continuous updating to reflect new model characteristics and skills, including increased evaluation awareness (Claude Sonnet 4.5 System Card, 2025; Apollo Research, 2025; Schoen et al., 2025; Needham et al., 2025).
[21] For example, document automation is a low-stake setting.
[22] DoW and the IC should consider the effect of sharing classified information with potentially misaligned models during operational T&E, and address it by starting with settings that operate on unclassified information (e.g., IL2, IL4).

Through our second recommendation, DoW, ODNI, NIST, and CAISI can ensure that the upcoming guidance:

- Strengthens federal agencies' ability to appropriately and rigorously consider potential threats "from the outset of an AI project" and run appropriate testing and mitigation in a "continuous" manner across an AI model's entire life-cycle (AIM Initiative, 2019; DoD AI Ethical Principles, 2020; DoD Responsible AI Strategy and Implementation Pathway, 2020; DIU RAI Guidelines, 2021; Political Declaration on Responsible Military Use of AI and Autonomy, 2023; Framework to Advance AI Governance and Risk Management in National Security, 2024), and especially prior to deployment (AI Ethics Framework for the IC, 2020; NSA Guidance for Strengthening AI System Security, 2024; NSA, Best Practices for Deploying Secure and Resilient AI Systems, 2024).

- Accelerates developmental T&E to its maximum potential, relying on it to "identify and diagnose undesired behaviors" (such as scheming) early on in controlled environments and "characterize and assess means to disengage or deactivate systems when necessary" (such as control measures) (DoD DT&E Guidebook, 2025).[24] Developmental T&E is the most effective stage to perform the aforementioned suite of scheming and control evaluations and ensure that an AI model is sufficiently reliable and governable before it is tested and evaluated in realistic operational settings with access to and permission to use the relevant affordances.

## 2.3. *How*: Acceptable Failure Rates

In the previous two recommendations, we recommended *what* DoW and the IC should evaluate in order to operationalize the principles of AI model reliability and AI model governability (§2.1) and *when* they should optimally perform these evaluations during existing T&E pipelines (§2.2).

Next, we bring together our previous two recommendations, and concentrate on ***how*** DoW and the IC can make informed decisions on AI model reliability and AI model governability after performing the recommended evaluations. In other words, here we concentrate on how scheming evaluations and control evaluations can concretely help DoW and the IC decide whether an AI model is sufficiently reliable and governable, or not.

---

[23] For instance, testers within DoW and the IC could rely on weaker trusted monitors (Bhatt et al., 2025) to flag suspicious outputs of the AI model being tested operationally. If the monitor flags a suspicious output, testers could save that context window and, instead of continuing this context in production, revert to developmental T&E and run thousands of simulations to predict what further actions the AI model might have taken if it had had the opportunity to.

[24] Indeed, "[e]valuating the reliability of systems under test is a well-established role of DT&E" (DoD DT&E Guidebook, 2025).

> **Recommendation (3)**
>
> We recommend that, in the upcoming guidance, DoW, ODNI, NIST, and CAISI clarify that:
>
> (A)   *Before* commencing testing, federal agencies should:
>
> - Define **acceptable failure rates** for scheming evaluations and control evaluations (§2.1).[25] Acceptable failure rates should consider both the measure of scheming rates and the 'circumstances'[26] in which scheming behavior manifests.[27]
>
> - Establish a **minimum viable procedure for running evaluations**. These baselines should, at a minimum, enumerate the scheming and control evaluations that will be performed during developmental T&E and operational T&E (§2.1), and elaborate on necessary AI model access, time, resources, and number of tests.
>
> - Leverage Testing & Evaluation Master Plans (TEMP) for these purposes.[28]
>
> (B)   *After* testing, **compare** the evaluation results of the developmental T&E and operational T&E against (A) to determine whether an AI model is sufficiently reliable and governable.

With regard to the acceptable failure rates, we observe that the science of evaluations is not yet close to being able to identify a precise threshold for scheming. This means that, while the science of evaluation advances, acceptable failure rates should not be seen as a strict threshold but rather as a tool to identify what clearly falls outside of tolerable levels. In this respect, we recommend that acceptable failure rates should be expressed as a spectrum, tailored to the sensitivity of the use case (for instance, warfighting), and potentially overcorrected to avoid scheming.

Through these recommended clarifications, DoW, ODNI, NIST, and CAISI can:

---

[25]   For clarity, acceptable failure rates could also be 0%, depending on the deployment settings and the correlated risk profile. Some AI usages and applications—for instance, warfighting (DoW, 2025; DoW, 2025)—may require conservative acceptable failure rates.

[26]   Circumstances to consider are, for instance, whether evaluators instruct the AI model to pursue a misaligned goal and provide it with situational awareness, or vice versa whether evaluators do not give the AI model a misaligned goal and do not provide it with situational awareness in context.

[27]   For example, some failure rates may be considered acceptable if an AI model is provided with situational awareness, given a misaligned goal, and strongly nudged to pursue it, but can be completely unacceptable if an AI model develops misaligned goals and situational awareness independently based on its training data (i.e., out-of-context).

[28]   As an illustrative example, a TEMP could require that: 'in [*x*] thousand hours of human red-teaming (by evaluators with [*x*] experience, [*x*] resources, and [*x*] model access), the developmental T&E team should not be able to find more than [*x*] instances of [e.g., no-goal] [e.g., in-context] [e.g., covert whistleblowing] over [*x*] thousand tests of type [*x*].'

- Build on and refine existing IC assurance frameworks, which recommend "[c]reat[ing] the discipline to define metrics up front and establish[ing] rigorous testing regimes and schedules" (such as, for instance, the AIM Initiative, 2019).

- Leverage the full potential of TEMP, which is the optimal T&E instrument to preemptively lay out the testing program and metrics, as well as objectives and thresholds (DoDI 5000.89, 2020; DoDI 5000.87, 2020; DoDM 5000.100, 2024).

- Enable DoW and the IC to make informed decisions not only on AI model deployment, but also on the transition of an AI model from developmental T&E to operational T&E. If developmental T&E shows scheming rates higher than the acceptable failure rates set in the TEMP, agencies should consider whether proceeding to operational T&E in combat-realistic conditions.

## 3. Interim Recommendations: Success Metrics in OT Agreements

In §2 above, our memorandum advanced recommendations to support DoW, ODNI, NIST, and CAISI in designing new guidance as requested by the AI Action Plan.

In this section, we address how DoW and the IC could improve and rapidly update the testing and evaluation of AI model reliability and AI model governability while the enactment of new guidance is still pending. It is important that AI models acquired during a period where new guidance is pending meet the same standards of reliability and governability as AI models procured after such guidance is enacted. AI models acquired by DoW and the IC while new guidance is pending could already have access to sensitive information by virtue of being "fine-tuned" on DoW's own data and operating in "classified environments" (Anthropic, 2025; Anthropic, 2025). Also, these AI models will likely never be publicly released. This entails that they will not undergo any sort of specialized external oversight prior to public release (e.g., OpenAI's o3 and o1, or Anthropic's Opus 4) and that the AI research community will be unable to vet their security post release.

We propose that, starting from today, DoW and DHS should **incorporate in their OT agreements** (10 U.S. Code § 4021; 6 U.S.C.A. § 391) **provisions that strengthen the principles of AI model reliability and AI model governability**.[29] In this respect, we see prototype OT agreements as a particularly high-leverage instrument for DoW and DHS. If successfully completed,[30] prototype OT agreements can lead to follow-on production OT agreement awards (10 U.S. Code § 4022(f)). Prototype OT agreements are considered "complete" when an approving official determines that the "**success metrics**" incorporated into the prototype OT agreement are satisfied (Secretary of Defense, 2018; DARPA, 2025). Success metrics can be a combination of objective and subjective measures (Under Secretary of Defense for Acquisition and Sustainment, 2023). Therefore, our final recommendation builds on the opportunity for DoW and DHS to embed expectations around scheming evaluations and control evaluations (§2.1) within these "success metrics."

---

[29] In this memorandum, we concentrate on DoW and the IC because the AI Action Plan tasks them to refine existing AI assurance frameworks. However, we note that similar recommendations could apply to the Department of Energy, which also has OT authority (42 U.S.C. 7256(g); 2 CFR Part 930).

[30] Successful completion is achieved if the prototype OT agreement "[m]et the key technical goals of a project" and "satisfied success metrics incorporated into the Prototype OT" (Secretary of Defense, 2018).

> **Recommendation (4)**
>
> We recommend that, pending the enactment of the upcoming guidance pursuant to the AI Action Plan, DoW and DHS operationalize the principles of AI model reliability and AI model governability through their OT authority.
>
> Specifically, we recommend that, when entering into prototype OT agreements, DoW and DHS:
>
> (A) Leverage prototype OT agreements' **success metrics** to:
>
> - Embed expectations on scheming evaluations and control evaluations (§2.1), in the form of acceptable failure rates (§2.3).
>
> - Establish a minimum viable procedure for running evaluations, including by elaborating on necessary AI model access, time, resources, and number of tests (§2.3).
>
> (B) **Subordinate the award of production OT agreements** to conditions of successful completion that incorporate these success metrics.[31]

As the science of evaluation advances, DoW and DHS could also leverage OT agreements to rapidly improve and update the operationalization of the principles of AI model reliability and AI model governability. For instance, in the future, DoW and DHS could rely on OT agreements' flexibility to require prototypes of 'AI safety cases' from developers.[32]

## 4. Conclusion

The open scientific problem of misalignment and its implications on AI model behavior, including scheming, cast a shadow on the principles of AI model reliability and AI model governability, both underpinning existing assurance frameworks by DoW and the IC. A misaligned AI model that acts on its scheming capabilities may adopt undesired behaviors that conflict with a government user's goals, and resist user's attempts to disengage or deactivate the model. For instance, if deployed in classified environments and given access to all interfacing systems, a misaligned AI model could deliberately lie, covertly whistleblow confidential data to the press or staff without security clearance, display dangerous capabilities that it had strategically hidden during evaluations (for instance, cyber or CBRN capabilities), or attempt to self-exfiltrate or blackmail government users to resist shutdown. In this sense, misaligned and scheming behavior in an AI model presents a red flag indicating that a model might be insufficiently reliable or governable and should be rigorously evaluated and controlled for.

---

[31] For instance, if an AI model custom-built pursuant to a prototype OT agreement shows measures of scheming rates higher than what defined within the success metrics, DoW and DHS should not award a follow-on production OT agreement.

[32] AI safety cases are still an emerging science. They are theorized as structured, evidence-based rationales that an AI system deployed in a specific setting is unlikely to cause unacceptable outcomes (Clymer et al., 2024; Buhl et al., 2024; Balesni et al., 2024; Goemans et al., 2024; Korbak et al., 2025). The goal would be requiring AI developers to provide DoW and DHS with structured, evidence-based rationales that the probability of the custom-built AI models (i) being misaligned and (ii) causing harm if misaligned are both sufficiently low.

This policy memorandum put forward three recommendations to support DoW and the IC in addressing the implications of misalignment and scheming on the principles of AI model reliability and AI model governability. Specifically, this memorandum recommends that, in the upcoming guidance, DoW, ODNI, NIST, and CAISI direct federal agencies to:

(1) Perform a suite of scheming evaluations and control evaluations in order to assess whether an AI model shows 'red flags' of insufficient reliability and governability.

(2) Perform these evaluations in controlled environments during developmental T&E and repeat them with near-production data in operational T&E, starting with low-stakes settings and unclassified information and moving toward higher-stakes settings in an iterative and incremental manner.

(3) Make informed decisions on AI model reliability and AI model governability by pre-defining acceptable failure rates and evaluation baselines (for instance, in the TEMP) and comparing the results of developmental T&E and operational T&E with these expectations.

In addition to advancing recommendations for the upcoming guidance that DoW, ODNI, NIST, and CAISI are tasked to prepare under the AI Action Plan, this memorandum provides a final suggestion on what DoW and the IC could do today to strengthen the principles of AI model reliability and AI model governability. We recommend that DoW and DHS:

(4) Leverage their OT authority strategically, and embed expectations on scheming and control evaluations within prototype OT agreements' success metrics, in the form of acceptable failure rates. This would provide agencies with a high-leverage instrument to deny follow-on production OT agreements if evaluation results of either scheming or control were unsatisfactory.